\documentclass[final,5p,times,twocolumn]{elsarticle}

\usepackage{lipsum}
\usepackage[pscoord]{eso-pic}

\newcommand{\placetextbox}[3]{
  \setbox0=\hbox{#3}
  \AddToShipoutPictureFG*{
    \put(\LenToUnit{#1\paperwidth},\LenToUnit{#2\paperheight}){\vtop{{\null}\makebox[0pt][c]{#3}}}%
  }%
}%

\usepackage{graphicx}

\usepackage{amssymb}

\usepackage{lineno}

\usepackage{hyperref}

\journal{Nuclear Instruments and Methods in Physics Research A }

\begin{document}

\begin{frontmatter}



\title{Experimental and Theoretical Progress of Linear Collider Final Focus Design and ATF2 Facility}


\author[label1]{Andrei Seryi}   \ead{andrei.seryi@adams-institute.ac.uk}
\author[label2]{Rogelio Tomas}
\author[label2]{Frank Zimmermann}
\author[label3]{Kiyoshi Kubo}
\author[label3]{Shigeru Kuroda}
\author[label3]{Toshiyuki Okugi}
\author[label3]{Toshiaki Tauchi}
\author[label3]{Nobuhiro Terunuma}
\author[label3]{Junji Urakawa}
\author[label4]{Glen White}
\author[label4]{Mark Woodley}
\author[label5]{Deepa Angal-Kalinin}
\address[label1]{John Adams Institute for Accelerator Science at University of Oxford, UK}
\address[label2]{CERN, Switzerland}
\address[label3]{KEK, Japan}
\address[label4]{SLAC, USA}
\address[label5]{STFC, ASTeC, UK}


\begin{abstract}
In this brief overview we will reflect on the process of the design of the linear collider (LC) final focus (FF) optics, and will also describe the theoretical and experimental efforts on design and practical realisation of a prototype of the LC FF optics implemented in the ATF2 facility at KEK, Japan, presently being commissioned and operated.   
\end{abstract}

\begin{keyword}
linear accelerator
\sep linear collider
\sep final focus
\sep beam optics
\sep chromaticity
\sep aberrations


\end{keyword}

\end{frontmatter}

\placetextbox{0.5}{0.06}{\fbox{{Published in Nuclear Inst. and Methods in Physics Research A 740 (2014) 2-5 ~  \url{http://dx.doi.org/10.1016/j.nima.2013.12.029}}}}%



\section{Introduction}
\label{sec-intro}



The future Linear Collider will require focusing the beams to nanometere sizes at the Interaction Point (IP) in order to reach desired luminosity \cite{ilc-tdr}, \cite{clic-cdr}. Production of the beams with ultra-low emittances and preservation of their quality is of paramount importance and have been the focus of both theoretical and experimental investigations by the linear collider community.

The ATF test facility has been constructed at KEK, Japan, in the end of eighties - early nineties in order to perform linear collider studies primarily focusing on generation of low emittance beams. The ATF facility, located in the hall situated inside of the perimeter of the TRISTAN ring (former, now Super-KEK-B ring), consists of an 1.3~GeV S-Band linac and Damping Ring (DR). The beam can be extracted from the DR to a short extraction line, equipped with various beam diagnostics. Detailed description of ATF facility can be found in \cite{atf-original}. 

The ATF facility has demonstrated achieving beam emittances suitable for linear collider, in particular the normalized emittances reached $\gamma\varepsilon_x = 3-4 \times 10^{-6}$~$m\cdot rad$ and $\gamma\varepsilon_y = 1-1.5 \times 10^{-8}$~$m\cdot rad$ in the range of the beam population of $0.1 - 1.0 \times 10^{10}$ \cite{atf-low-e}. 

Following and in parallel with studies of the ultra-low emittance beam, various other studies have been conducted at ATF facilities from 1997 to 2008, in particular studies of multi-bunch acceleration and damping (up to 20 bunches per train, up to 3 trains per DR cycle), studies of ultra fast kicker with rise time less than 3~ns, development of micrometer resolution laser wire beam diagnostics, development of laser cavity and Compton laser-beam interaction, development of Fast Feedback on Nanosecond Time Scale (FONT), development of nanometer resolution cavity beam position monitors, studies of fast ion beam instabilities, etc. Comprehensive information on ATF studies can be found in \cite{atf-web}.    

Preparations for the realisation of the linear collider required experimental studies of the final focus system. The ATF facility was found to be best suited for FF experimental studies. Factors that played key role in selection of ATF as the test bed for FF studies, were the following: a) beams with beam emittances compatible with linear collider requirements; b) state of the art beam facility well equipped with beam diagnostics and dedicated for LC studies; c) active international collaboration working on ATF; d) infrastructure compatible with expansion of the facility; e) support of the host laboratory -- KEK.  

Before we will discuss the ATF2 facility, a prototype of ILC final focus system realised at KEK, let us recall the different approaches to LC final focus design and in particular discuss why was it necessary to perform experimental study of a FF, despite the fact that the first experimental demonstration of a linear collider final focus system has been done at FFTB facility at SLAC in early nineties \cite{fftb} where focusing of the beam down to approximately 70~nm has been achieved, and compensation of chromaticity by pairs of non-interleaved sextupoles placed in the dedicated optical sections upstream of the IP has been demonstrated successfully.

\section{Requirements to FF and to BDS}
\label{sec-ff-reqs}

The Final Focus is a part of a so called Beam Delivery System (BDS) of a linear collider, which also includes the energy and betatron collimation systems, diagnostics section, main extraction line, tune-up extraction line, beam dumps, precise energy spectrometer and polarimeter, machine-detector interface (MDI) elements, etc \cite{ilc-tdr}.   

Some of the key factors that drives the design of BDS and in particular of FF are as follows. The large chromaticity created by Final Doublet (FD) requires careful cancellation of chromaticity and other associated aberrations -- this will be the main topic of discussion further in this paper.  Strong beam-beam effects and associated background in the detector, due to primary beams as well as secondary particles, define the design of the near IP beam elements, detector components, and also affect the geometry of the beamlines and the value of the crossing angle at the IP. The need for minimization of emittance and energy spread growth due to synchrotron radiation in the bends in the FF and BDS, at the required near TeV energies, affects the length of the BDS and its scaling with beam energy, and forces the use of weak and long dipoles. The need to collimate the beam halo strongly affects the design and length of the collimation section, and together with the beam train format defines whether the collimation system spoilers and absorbers need to be survivable or consumable. Requirements for the beam diagnostics sections come, in particular, from the need to make the beam sizes sufficiently large in comparison with the practically achievable resolution of the laser wire beam size monitors. All these and other requirements have been taken into account in the design of ILC BDS. Let us concentrate now on the requirements coming from the chromaticity of the final doublet. 

In order to achieve the smallest beam size at the IP, the strongest lens in the FF has to be the final doublet, located as close as practical to the IP. Therefore the FD determines the chromaticity of the FF. In a thin lens approximation the chromatic dilution of the beam due to FD can be estimated as $\Delta\sigma/\sigma \sim \sigma_{\mathrm{E}} \mathrm{L}^* / \beta^*$ where $\sigma_{\mathrm{E}}$ is the beam relative energy spread (typical values are 0.002-0.01), $\mathrm{L}^*$ is the distance between IP and the FD (typically 3-5~m) and $\beta^*$ is the betatron function at the IP (typically 0.4 - 0.1~mm). For typical values given above the beam chromatic dilution would range from 15 to 500, which is enormous. Therefore the primary requirement to a FF is to focus the beam while compensating the beam chromaticity and any other important associated aberrations.

\section{Final Focus designs}
\label{sec-ff}

\begin{figure*}[htb!]
\centering
\includegraphics[width=0.98\textwidth]{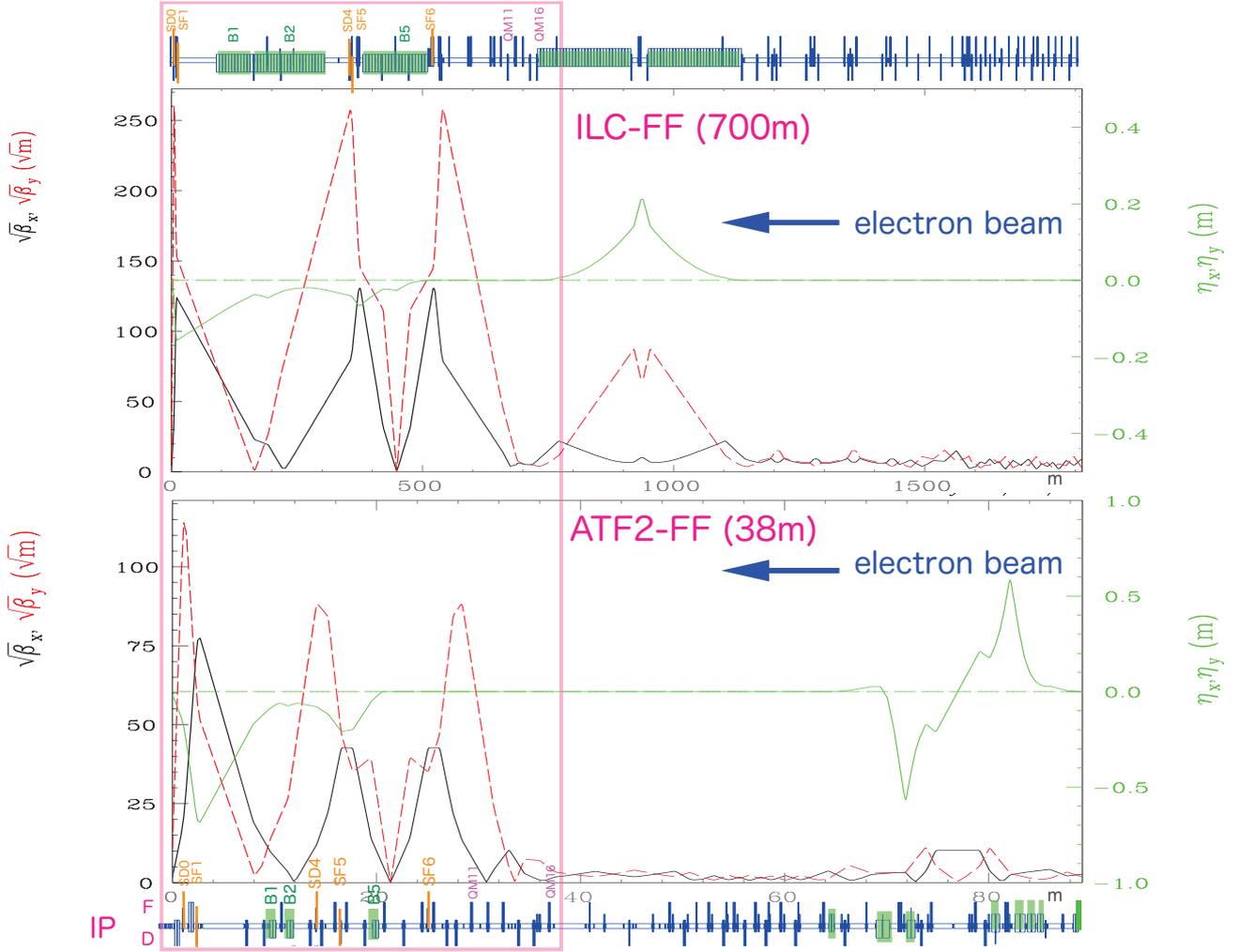}
\caption{Optics of ILC BDS (above) in comparison with optics of ATF2 FF \& extraction and diagnostics beamline.}
\label{fig:ilc-atf}
\end{figure*}

Historically, the first designs of the FF for linear collider contained four sections: the matching telescope, the horizontal chromaticity compensation section, the vertical chromaticity compensation section, and the final telescope. The chromaticity compensation sections consisted of symmetric optics which created two locations with large beta-functions in both planes as well as maxima of dispersion function, where sextupoles were placed. The transfer matrix between sextupoles were designed to be $\mathrm{M} = -\mathrm{I}$ in order to cancel the aberrations produced by sextupoles for on-energy particles, while creating additional focusing/defocusing effect for off-energy particles, to compensate the FD chromaticity. The two pairs of $-\mathrm{I}$ separated sextupoles, for x and y chtomaticity compensation, were typically non-interlaced, to minimize the third and higher order aberrations. All earlier designs followed this principle. Examples include one of the first LC FF design \cite{zotter,zotter-napoly}, the FFTB optics \cite{fftb-optics} later realized experimentally, the JLC FF optics \cite{jlc-ff}, the VLEPP linear collider FF optics \cite{vlepp-ff}, the NLC final focus \cite{nlc-ff}, etc.  The latter one, NLC FF designed for TeV energy reach, was approximately 1.8~km long and had  $\mathrm{L}^* = 2$~m.  

These first FF were later dubbed "traditional" or non-local chromaticity compensation FF. The issues of the traditional FF follow from the design -- the chromaticity is created in the FD near IP, but is pre-compensated $\sim 1000$~m upstream. Any disturbances to the beam, for example the synchrotron radiation generated energy spread, created between sextupoles and IP, would disturb compensation of the chromaticity. Compensation of sextupole aberrations in such a system is also not ideal as $\mathrm{M} \not= -\mathrm{I}$ for off-energy particles. This in particular creates large aberrations for off-energy particles and especially for the beam tails. 

Various improvements to the "traditional" design have been suggested and developed in the past. One can in particular note the suggestion to use additional sextupoles (so called Brinkmann's sextupoles) and adjusting dispersion function to increase the energy acceptance of a FF \cite{brinkmann-sextupoles}, and a suggestion to alter the behaviour of dispersion function to improve the control of aberrations and increase the FF bandwidth \cite{oide-odd-dispersion} (so called Oide's odd-dispersion scheme). These improvements, however, did not significantly improve the multi-TeV performance and scaling of a FF. 

A radically alternative FF design has been suggested later, featuring local compensation of chromaticity \cite{ff-prl}. In this design: a) the chromaticity is cancelled locally by two sextupoles interleaved with FD;  b) a bend upstream generates dispersion across the FD needed for sextupoles to work; c) the parasitic second order dispersion produced by sextupoles in the FD is cancelled locally provided that half of horizontal chromaticity arrives from upstream of FD; d) the geometric aberrations of the FD sextupoles are cancelled by two more sextupoles placed in phase with them and upstream of the bend; e) the higher order aberrations are cancelled by optimizing transport matrices between sextupoles. The design feature c) comes from the fact that two sextupoles placed in the FD cannot simultaneously cancel three parameters -- the X and Y chromaticity and the X-second order dispersion, however introducing a new free parameter, the amount of horizontal chromaticity arriving upstream of FD, allowed to cancel all three major lower order aberrations simultaneously. 

The first FF based on a local chromaticity compensation principle has been designed to work for NLC parameters and it was found that with the same performance it can be just $\sim$300~m long, i.e. 6 times shorter than the traditional design \cite{ff-prl}. Moreover, it was found that the FF required length scales only as $\mathrm{E}^{2/5}$ with energy which means that one can design a multi-TeV FF in under a km length. 

Such an improved performance in a short length allowed to relax certain constraints of the FF, and in particular to double the $\mathrm{L}^*$ -- the later versions of NLC and ILC FF were designed for $\mathrm{L}^*$ of 3.5 to 4.3~m which allowed significant simplification of the design of MDI elements. The bandwidth (energy acceptance) of FF with local chromaticity correction was found to be better than for a system with non-local correction. It was found that FF with local chromaticity correction has much less aberrations and it does not mix betatron phases of non-core particles, which is in particular important for minimization of generation of beam tails and ensuring good performance of the collimation system.  

The payback for the improved performance of the FF with local chromaticity compensation was much more difficult and quite lengthy process of its design, which was caused by the fact that good cancellation of higher order aberrations required optimal selection of the first order optics. Eventually, the procedure and recipe for FF design and optimisation have been found and tools assisting design and allowing control of aberrations up to fourth order have been developed, permitting to approach the design process semi-algorithmically and reducing the time of the optics development. The recipe and example of a system designed according to the developed procedure is given in \cite{ff-recipe}. 

The ATF2 Final Focus, which was conceived as a prototype of a FF with local chromaticity compensation, was designed using the same principles and procedures. More recently, advanced optics design and nonlinear optimisation tools based on MAPCLASS code have been developed \cite{tomas-nonlinear} and used for further optimisation of ATF2 optics as described further below.

\section{ATF2 Optics Design}
\label{sec-atf2-ff}

Given the fact that after invention of FF with local chromaticity compensation the FF of all linear collider designs were relying on this new approach, yet it was not investigated experimentally, the LC community initiated discussion of the idea to create a test facility where local chromaticity correction FF could be studied. 

The idea of a new test facility at ATF, to prototype the advanced final focus for linear collider, was first conceived in 2002 at Nanobeam workshop in Lausanne.  An extended version of ATF was presented \cite{junji-nanobeam2002} where ATF2 facility would include a new compact final focus and also a new X-band section of the linac (which was not realised). The FF optics for ATF2 at that time was based on a compact design \cite{kuroda-nanobeam2002}, yet not on local chromaticity compensation, and such design could not be scaled to TeV energies. Consequently, when the ATF2 idea started to gain momentum in 2005, the ATF2 FF optics has been changed to a scaled down version of ILC FF optics, re-designed and optimised according to local chromaticity compensation principles \cite{first-atf2-workshop-ff}. The final version of ATF2 optics, in comparison with ILC BDS optics, is shown in Fig.\ref{fig:ilc-atf}.

The next section briefly describes realization of ATF2. In parallel with realization of the baseline ATF2 design, the collaboration has been looking further, and in particular developed an ultra-low IP beta function version of ATF2 optics \cite{atf2-ultralowbeta} which is considered for realization in the future.

\section{ATF2 Realisation}
\label{sec-atf2-org}

Following the first ATF2 Workshop in 2005, the ATF international collaboration and the ATF2 project team organized itself to prepare an MOU for ATF2 realisation and the ATF2 Proposal which was signed by 110 authors from 25 institutions \cite{ATF2-vol1}. Early in 2006 the second volume of ATF2 Proposal has been published, which included the construction schedule and the cost information, as well as the planned in-kind contribution by collaboration partners \cite{ATF2-vol2}. 

Aiming to develop the experience that will be suitable for eventual realization of ILC, the ATF collaboration created an efficient organizational structure for ATF2 implementation. The ATF Spokesperson was assisted with three Deputies -- to carry out tasks in the areas of a) Beam operation; b) Hardware maintenance; c) Design, construction and commissioning of ATF2. As some of the deputies were international colleagues, local KEK sub-deputies were assisting in their duties. 

The ATF Spokesperson was responsible for directing and coordinating the work required at ATF/ATF2 in accordance with the ATF Annual Activity Plan, was reporting the progress to the International Collaboration Body (ICB) and the progress and the matters related to KEK budget to the KEK Director General. The ICB was set up to be the decision making body for executive matters related to the ATF collaboration.  A Technical Board (TB) was created to assist the Spokesperson in formulating the ATF Annual Activity Plan, including the budget and beamtime allocation and assist the ICB in assessing scientific progress. 

ATF2 has been constructed in ILC model, with in-kind contribution from partners and host country providing civil construction, and major part of operation support. 
The ground breaking for construction of ATF extension, to house the FF, started in 2007, when a new reinforced piled concrete floor and shielded area have been constructed. 

Examples of hardware contribution by ATF2 partners included in particular high availability power supplies, magnet movers and BPM electronics provided by SLAC, beamline magnets built by IHEP (China), cavity BPM system built and developed by collaboration of PAL (Korea) and JAI/RHUL (UK), the Final Doublet assembled by collaboration of SLAC and LAPP teams, etc. One of the most important instruments which allowed measuring the focused beam size at the IP was the interferometric IP Beam Size Monitor (IPBSM) developed by the University of Tokyo team described in details in \cite{IPBSM-EAAC} .

\section{ATF2 Commissioning and recent results}
\label{sec-atf2-comm}

The ATF2 commissioning and first results have been described elsewhere. In particular, the paper \cite{atf2-prstab-2010} described the status of ATF2 as of 2010 when most of the beam hardware been commissioned and regular operation with the beam started.

The most recent commissioning and beam operation lessons have been described in the ICFA Beam Dynamics Newsletter dedicated to ATF2 \cite{ATF2-BD-Newsline-2013}. 

The most recent experimental results are being prepared for publication \cite{to-be-submitted}. In the present paper, we would like to stress that sufficiently small beam sizes have been achieved at ATF2 allowing to confirm successful operation of ATF2 as the prototype of LC FF, and the developed FF tuning methods and instrumentation confirmed the design principles and to a large extent the expected performance of the ILC Final Focus.

\section{Conclusion}
\label{sec-conclus}

The Final Focus with local chromatic correction works in theory and in practice. The ATF/ATF2 international collaboration successfully demonstrated operation of ILC-like final focus system. The ATF2 project was realized as ILC-like international project, with in-kind contributions from many international partners. The ATF2 is a great training and advanced accelerator research facility with a great potential to serve the needs of the accelerator community.



\bibliographystyle{elsarticle-num}
\bibliography{<your-bib-database>}



\end{document}